\documentclass[prl,nofootinbib,showpacs,showkeys,twocolumn]{revtex4}
\usepackage{exscale}
\usepackage{amssymb,amsbsy}
\usepackage[dvips]{graphicx}
\usepackage{graphicx}
\usepackage{dcolumn}
\usepackage{bm}
\usepackage{amsmath}
\usepackage{epsf,epsfig,graphicx}

\begin{document}
\title{ Momentum Correlations for Identical Fermions }
\author{Z.~Yang$^{a,b}$}
\author{X.~Sun$^c$}
\author{Y.~Gao$^{a,b}$}
\author{S.~Chen$^{a,b}$}
\author{J.~Cheng~$^{a,b}$}
 \affiliation{$^a$ Department of Engineering Physics, Tsinghua
University, Beijing 100084, China\\
$^b$Center for High Energy Physics, Tsinghua University, Beijing
100084, China\\
$^c$ Institute of High Energy Physics, Beijing 100039, China}

\begin{abstract}
\setlength{\baselineskip}{16pt}
 The effects of intensity
interferometry (HBT), which are mainly reflected by the measured
momentum correlations, have been fully discussed since 1950's. The
analogous momentum correlations between identical fermions are
however less argued. Besides the fermion statistics, the momentum
correlations between identical fermions are also determined by some
other aspects, such as coulomb interaction, spin interactions and
resonances formation. In this paper, we discussed the factors that
influence the momentum correlations between $\Lambda$'s, especially
spin interaction. It is also argued that the momentum correlations
between $\Lambda$'s are affected significantly by the yield ratio of
$\Sigma^0/\Lambda$, due to the limitation of experimental
measurement. The mixture of $\Sigma^0$'s and $\Lambda$s would
significantly weaken the measured two-particle momentum correlations
of $\Lambda$'s.
\end{abstract}

\maketitle

\section {Introduction}

Intensity interferometry(HBT), which was firstly proposed by
Hanbury-Brown and Twiss\cite{hanbury-brown56} as a method of
measuring the sizes of stars, was later extended to particle physics
by Goldhaber \emph{et al}\cite{goldhaber60} to extract the
space-time structure of $p\bar{p}$ annihilations by studying the
angular distribution of identical pion pairs in the collisions. HBT
interferometry is now regarded as a useful tool to study the
space-time structures of fireballs produced in heavy-ion
collisions\cite{heinz99,pratt84}. The HBT interferometry is an
effect of Bose-Einstein statistics, which leads to an enhancement of
the identical boson pairs when their relative momenta is small. The
effect is usually represented by the measured two-particle momentum
correlations. It has been argued that some dynamical information of
the expanding fireball created in heavy-ion collisions can be
obtained by the correlations. For instance, the information on the
anisotropic transverse flow of the fireball can be extracted by
measuring the correlation functions as a function of the emission
angle\cite{voloshin96_1,voloshin96_2,heinz02}. In fact, the
2-particle momentum correlations of a many-body system are
determined not only by the quantum statistics, but also the
interactions between the particles produced in the source and the
final-state interactions. The pure quantum statistics effects can be
acquired only after all these influences have been removed
carefully. For example, due to the repulsive coulomb interactions,
the momentum correlations between identical charged pions are
suppressed to some extent for small relative momenta, and the
expected great enhancement can show up only after the coulomb
interactions are removed correctly\cite{star01_1}.

Although the theory of HBT effects\cite{heinz99,pratt84,csorto05}
has been fully discussed since 1950's, and a lot of experimental
results have been aquired by relativistic heavy-ion collisions at
CERN/SPS\cite{wa9799,appelshauser99} and
RHIC\cite{star01_1,phenix02,phenix03_1,star03_2,star03_3,star04_1,star04_2},
there are still a lot of open questions in this area. Both theorists
and experimentists are embarrassed by the socalled ``HBT
puzzle"\cite{magestro05} that hydrodynamic calculations, which
provide perfect results of anisophic flows, yield strong
disagreement with HBT radii and predict lack of energy dependence of
the HBT radii for a fixed $k_T$ bin\cite{heinz02_hydro,kolb03}.
Besides ``HBT" puzzle, the analogous momentum correlations between
identical fermions are less argued. The conclusion can be made
intuitively that the identical fermion pairs should be suppressed
for small relative momentum due to Fermi-Dirac statistics. But
little knowledge has been obtained by now to make a definite
conclusion. Different from Boson's correlations, the spin
interactions play an important role in the two-particle momentum
correlations between identical fermions. The abundance of protons
produced in heavy-ion collisions guarantees the statistics for
momentum correlations between them, and SPS did measure the
correlations \cite{appelshauser99}. But the complicated interactions
between protons and the strong coulomb interactions are always big
issues, which lead to the difficult to draw a definite conclusion.
C. Greiner and B. M\"uller\cite{Greiner89} proposed the momentum
correlations between $\Lambda$'s and discussed the influences  by
the possible existing di-$\Lambda$(H
particle)\cite{jaffe77a,jaffe77b}. WA97 measured the
$\Lambda$-$\Lambda$ correlations\cite{wa9799} in Pb-Pb collisions at
158A GeV/c, but poor statistics refused any expected conclusions.

The relativistic heavy-ion collisions at RHIC is a factory of
strangeness, where quite a lot of $\Lambda$'s are produced for
central collisions. This gives us a chance to measure the momentum
correlations between $\Lambda$'s with better statistics. The most
important advantage of the analysis of $\Lambda$ correlations is
that the coulomb interactions, if exist, can be neglected. What's
more, We hardly had any opportunity before to study the
interactions between baryons with strangeness, the momentum
correlations between $\Lambda$'s may give us a first glimpse of
the interactions of strange baryons, however rough it might be.
The investigation of the momentum correlations between $\Lambda$'s
at RHIC is therefore interesting and valuable at least from this
point of view.

In this paper, we discussed those important factors that influence
the momentum correlation between $\Lambda$'s , one of which is the
spin interactions between $\Lambda$'s. It is also argued that the
measured momentum correlations between $\Lambda$'s are affected
significantly by the yield ratio of $\Sigma^0/\Lambda$ due to the
limitation of experimental measurement.

The outline of this paper is as follows. The momentum correlation
functions are presented in section 2, and the effects of spin
interactions are argued in section 3. In section 4 we discussed
the effects of $\Sigma^0/\Lambda$ ratio on the measured momentum
correlations. Finally a brief summary is given in the last
section.

\section{The 2-particle Correlation Function}
The two-particle correlation function $C_2(\vec p_{rel})$ is
usually defined as
\begin{eqnarray}
 \label{correlation1}
  C_2(\vec p_{rel})&=&\frac{P(\vec p_1,\vec p_2)}{P(\vec p_1)P(\vec p_2)}\equiv 1+R(\vec p_{rel}),\\
  P(\vec p_1,\vec p_2)&=& \frac{1}{\sigma_{12}}\frac{d^6\sigma_{12}}{d^3\vec p_1d^3\vec p_2},\nonumber\\
  P(\vec p_i)&=& \frac{1}{\sigma_1}\frac{d^3\sigma_i}{d^3\vec p_i},\ (i=1,2),\nonumber
\end{eqnarray}
where $\vec p_{rel}=\vec p_2-\vec p_1$ is the relative momentum,
$P(\vec p_i$ represents the probability of detecting a particle
with momentum $\vec p_i$, and $P(\vec p_1,\vec p_2)$ that of
detecting two particles that one with momentum $\vec p_1$ and the
other with momentum $\vec p_2$. $R(\vec p_{rel})$ measures the
difference between the two-particle cross section and the product
of the inclusive single-particle cross section. Correlations
between particles would lead $R(\vec p_{rel})$ to deviate from
zero. If there are no correlations, $R(\vec p_{rel})$ will vanish.
Therefore $R(\vec p_{rel})$ is sometimes also called correlation
function. For thermalized expanding fireballs which emit identical
particles, the correlation function $C_2$ should depend on the
space size of the source at freeze-out.

For $\Lambda$-$\Lambda$ correlations, we can approximately adopt
the non-relativistic expression to calculate the correlation
function $C_2(\vec p_{rel})$. Following the steps by Greiner and
M\"uller\cite{Greiner89}, let $D(\vec r,t;\vec p)$ be the relative
probability that a particle with momentum $\vec p$ freezes out at
a place $\vec r$ at time $t$, and normalized on the differential
cross section,
\begin{eqnarray}
 \int d^3\vec rdt D(\vec r,t;\vec p)=\frac{1}{\sigma}\frac{d^3\sigma}{d^3\vec p}.\nonumber
\end{eqnarray}
The two-particle differential cross section may then be
approximated as
\begin{eqnarray}
 \label{dcrosssection2}
    & & \frac{1}{\sigma_{12}}\frac{d^6\sigma_{12}}{d^3\vec p_1d^3\vec p_2}\nonumber\\
    &=& \int_{-\infty}^{+\infty}dt_1dt_2\int d^3
        \vec r_1d^3\vec r_2D(\vec r_1,t_1;\vec p_1)D(\vec r_2,t_2;\vec p_2)\nonumber\\
    & & \cdot
        (2\pi)^6|\Psi_{1,2}({\vec r_1'},\vec r_2;\vec p_1,\vec p_2)|^2,
\end{eqnarray}
where ${\vec r}_1'=\vec r_1+\vec v(t_2-t_1)$, $ \vec v=\frac{\vec
p_1+\vec p_2}{2m}=\vec P_{cm}/2m$, and  $\psi_{1,2}$ is the exact
2-particle  wave function in the final channel. We parameterize
the distribution function $D(\vec r,t;\vec p)$ by the following
simple gaussian ansatz,
\begin{equation}
 \label{distribution1}
  D(\vec r,t;\vec p)
 =\frac{1}{\sigma}\frac{d^3\sigma}{d^3\vec p}\left(\frac{1}{\pi^{3/2}}\frac{1}{r_0^3}\exp(-r^2/r_0^2)\right)
         \delta(t-t_0),
\end{equation}
where $t_0$ denotes the moment of freeze-out, and $r_0$ defines
the spatial size of the fireball at freeze-out. The $\delta$
function indicates the ignorance of the formation time of
particles. Inserting (\ref{distribution1}) in
(\ref{dcrosssection2}) and comparing with (\ref{correlation1}), we
obtain
\begin{eqnarray}
 \label{correlationtmp}
  C_2(\vec p_{rel})
    &=& \frac{1}{\pi^3}\frac{1}{r_0^6}
        \int d^3\vec r_1d^3\vec
        r_2e^{-(r_1^2+r_2^2)/r_0^2}\nonumber\\
   && \cdot(2\pi\hbar)^6|\Psi_{1,2}(\vec r_1,\vec r_2;\vec p_1,\vec p_2)|^2.
\end{eqnarray}

\section{Effects of Spin Wave Functions}
The key step to calculate the correlation function
(\ref{correlationtmp}) is then to find the exact two-particle wave
function, $\Psi_{1,2}$. For baryons with strangeness, the
interactions between them is not yet clear, but the final-state
interactions of strange particles can be ignored as a first step,
and we can focus our attention on the effects of the underlying
quantum statistics, and express $\Psi_{1,2}$ simply as the
production of plane waves of the particles.

For bosons, the total wave function $\Psi_{1,2}$ is just the
spatial wave function itself, which should be symmetric under
particle exchanges. It is easy to show that
\begin{eqnarray}
 \label{correlation2bosons}
  C_2^B(\vec p_{rel})=1+\exp(-p_{rel}^2r_0^2/2).
\end{eqnarray}

The case becomes a little complicated for fermions, which demand
the total wave function to be asymmetric under particle exchanges.
The total wave is the product of spatial wave function and spin
wave function, and other internal wave functions such as isospin
wave function, if exist. For spin-$\frac{1}{2}$ particles, there
are spin singlet state, $\chi_A$, and triplet states, $\chi_S$,
for a 2-particle system. The former corresponds to spin asymmetric
state, while the latter spin symmetric state. We can thus form two
kinds of asymmetric total wave functions fermions with no other
internal space,
\begin{eqnarray}
 &&\psi^S(\vec p_1,\vec r_1;\vec p_2,\vec r_2)=\phi^+\chi_A,\nonumber\\
 &&\psi^T(\vec p_1,\vec r_1;\vec p_2,\vec r_2)=\phi^-\chi_S,\nonumber
\end{eqnarray}
where $\phi^+$ and $\phi^-$ are respectively normalized symmetric
and asymmetric spatial wave functions.
 Therefore, the total wave function, $\Psi_{1,2}$, can be
express as a mixed state of singlet $\psi^S$ and triplet $\psi^T$,
\begin{equation}
 \Psi_{1,2}=\alpha\psi^S+\beta\psi^T, |\alpha|^2+|\beta|^2=1.
\end{equation}
The momentum correlation function of $\Lambda$'s is then given by
\begin{eqnarray}
    C_2^\Lambda(\vec p_{rel}) =
    1-(|\beta|^2-|\alpha|^2)\exp(-p_{rel}^2r_0^2/2),
\end{eqnarray}
which depends apparently on the probability of spin singlet and
triplet stats. If we assume spin singlet state and each component
of triplet state has the same probability, that is, $\alpha=1/2$
and $\beta=\sqrt3/2$, the correlation function becomes
\begin{eqnarray}
  C_2^\Lambda(\vec p_{rel}) =
  1-\frac{1}{2}\exp(-p_{rel}^2r_0^2/2).
\end{eqnarray}
This is exactly the formula given by Greiner and
M\"uller\cite{Greiner89}.

Unfortunately we have in fact little (if not blank) knowledge
about the spin interactions between hyperons. It is hardly to
determine the value of $\alpha$ and $\beta$. The overall
investigation on the interactions between nucleons are made mostly
via $p$-$p$ collisions. Deuteron, the two-particle system of
$n$-$p$, is the only boundary state of nucleon-nucleon. The spin
of deuteron is $S=1$, which indicates that the spins of neutron
and proton intend to be in the same direction and form a triplet
state. But we have no idea whether a $\Lambda$-$\Lambda$ system
prefers triplet or singlet state.

If $\beta=\alpha=1/\sqrt2$, the correlation function $R_\Lambda$
would vanish, which means no effects of momentum correlations
exhibit. When $\beta>\alpha$, it gives $R_\Lambda<0$ and
Fermi-Dirac suppression would dominate. If $\beta<\alpha$, it
gives a desperate result, $R_\Lambda>0$, which means that
Bose-Einstein enhancement shows up!

If $\beta=1$, that is, the system is in pure triplet state,
$R_\Lambda=-\exp(-p_{rel}^2r_0^2/2)$, which represents a full
Fermi-Dirac suppression for small relative momentum. On the
contrary, if $\alpha=1$, that is, the system is in pure singlet
state, $R_\Lambda=\exp(-p_{rel}^2r_0^2/2)$, which represents that
the correlation behaviour of $\Lambda$'s is totally in the same
way as bosons.

\section{The Effects of $\Sigma^0/\Lambda$ Ratio}

The production of $\Sigma$'s in heavy-ion collisions gives
sensitive influence on the measured $\Lambda$-$\Lambda$
correlation functions. $\Sigma^0$ has the same value quark
components as $\Lambda$, but a little heavier. It decays to
$\Lambda$ and a low energy photon. With the mean life
$\tau\sim10^{-20}s$, which is far longer than the life of the
fireballs created by heavy-ion collisions, nearly all $\Sigma^0$'s
decay long after the freeze-out of the fireballs, compared with
the life of the fireballs. The decay length of $\Sigma^0$ is of
the order $10^4$fm, which is small enough to prevent us
distinguishing experimentally the decay vertex of $\Sigma^0$
between primary vertex, we can therefore hardly determine whether
a $\Lambda$, when reconstructed, is emitted directly from the
fireball or from $\Sigma^0$ decay.

Suppose that $N$ $\Lambda$'s are reconstructed in one event, which
are ``all" selected to be emitted from the source. If the feed-down
from multiply strange hyperons, notably $\Xi^0$ and $\Xi^-$, is
neglected, these $N$ $\Lambda$'s mainly come from two kinds of
contributions,``primordial" $\Lambda$'s that are really originated
directly from the fireball, and ``decayed" $\Lambda$'s that come
from $\Sigma^0$ decays.  Suppose that the yield ratio of
$\Sigma^0/\Lambda=\gamma$, typically $\gamma\leq1$. The number of
``primordial" $\Lambda$'s is then $n_{prim}=N/(1+\gamma)$, while the
number of ``decayed" $\Lambda$'s is $n_{decay}=\gamma N/(1+\gamma)$.
Since the ``decayed" $\Lambda$'s are nearly all produced chaotically
long after freeze-out of the fireball, there should be no
significant momentum correlations between them. There should also be
no harm to declare that there are no momentum correlations between
``decayed" $\Lambda$'s and ``primordial" $\Lambda$'s.

The correlation function $C_2=1+R$ is experimentally obtained by the
pairs in same events divided by pairs in mixed events. The pairs
obtained from the $N$ $\Lambda$'s can be separated into three parts,
those between ``primordial" $\Lambda$'s, $N(\Lambda\Lambda)$, those
between ``decayed" $\Lambda$'s, $N({\Sigma^0\Sigma^0})$ and those
between ``primordial" and ``decayed" $\Lambda$'s $,
N(\Lambda\Sigma^0)$,
\begin{eqnarray}
  &&N(\Lambda\Lambda)=\frac{1}{2}\frac{N}{1+\gamma}\left(\frac{N}{1+\gamma}-1\right),\nonumber\\
  &&N(\Sigma^0\Sigma^0)=\frac{1}{2}\frac{\gamma N}{1+\gamma}\left(\frac{\gamma N}{1+\gamma}-1\right),\nonumber\\
  &&N(\Lambda\Sigma^0)=\frac{\gamma N^2}{(1+\gamma)^2}.\nonumber
\end{eqnarray}
Totally,
$N(\Lambda\Lambda)+N(\Sigma^0\Sigma^0)+N(\Lambda\Sigma^0)=\frac{1}{2}N(N-1)=N_{pair}^t$,
where $N_{pair}^t$ is the total pairs of the $N$ $\Lambda$'s.

According to the discussion above, there is no contribution to the
correlation function for $N(\Sigma^0\Sigma^0)$ and
$N(\Lambda\Sigma^0)$, that is, $C_2=1$ for these two kinds of pairs.
The experimentally obtained correlation function is therefore
written as
\begin{eqnarray}
  C_2^{exp} &=& \frac{N(\Lambda\Lambda)}{N_{pair}^t}(1+R_\Lambda)
                +\frac{N(\Sigma^0\Sigma^0)}{N_{pair}^t}+\frac{N(\Lambda\Sigma^0)}{N_{pair}^t},\nonumber\\
            &=& 1+\frac{N-(1+\gamma)}{(1+\gamma)^2(N-1)}R_\Lambda,
\end{eqnarray}
where $R_\Lambda$ is the correlation function of pure $\Lambda$'s.
For $N\gg1$, $C_2^{exp}\simeq1+\frac{1}{(1+\gamma)^2}R_\Lambda$.
If $\gamma=1$, and suppose that the probability of singlet and
 each component of triplet is equal,
\begin{eqnarray}
  C_2^{exp} &=&
  1+\frac{1}{4}R_\Lambda=1-\frac{1}{8}\exp(-p_{rel}^2r_0^2/2).
\end{eqnarray}
This means the the effects of pure $\Lambda$'s momentum
correlations would be significantly weakened if we can not
identify those $\Lambda$'s from $\Sigma^0$ decay. So we can hardly
observe significant suppression of $\Lambda$-$\Lambda$
correlations in heavy-ion collisions.

\section{Conclusions}

we investigated the momentum correlation between identical
fermions, especially that between $\Lambda$'s produced in
heavy-ion collisions. Different from bosons, spin interactions
play an important role in the behaviour of fermion's correlations.
For $\Lambda$'s, if the spin wave function of $\Lambda$-$\Lambda$
system is symmetric, the asymmetric space wave function leads to a
total suppression at small relative momentum. If the spin wave
function is asymmetric, the symmetric space wave function,
however, leads to a total enhancement at small relative momentum,
just the same as the behaviour of identical bosons. The lack of
knowledge on the spin interactions makes it complicated to draw a
conclusion whether the correlation functions should suppress or
enhance at small relative momentum. It is also hard to say to what
extent the suppression or enhancement would be.

It is also argued that $\Sigma^0/\Lambda$ yield ratio affects the
measured momentum correlations of $\Lambda$. A high ratio would
weaken the the correlation effects significantly. For
$\Sigma^0/\Lambda=1$, the measured effect of correlations would be
reduced to only about one quarter of that of correlations between
pure $\Lambda$'s. In fact, the number of $\Lambda$'s $N$ that can be
reconstructed for Au-Au central collisions at RHIC\cite{star02} is
quite small. In this case, the deviation of $C_2^{exp}$ from $1$ is
therefore even smaller.

The trouble is that neither the spin interactions between
$\Lambda$'s nor the $\Sigma^0/\Lambda$ yield ratio is clear to us.
If the spin interactions could be clearly understood, the behaviors
of pure $\Lambda$'s correlations would be determined to a great
extent. The $\Sigma^0/\Lambda$ ratio would therefore be extracted
from the measured momentum correlation function. The
$\Sigma^0/\Lambda$ is important to to heavy-ion collision since it
can give us information on the mechanisms of hadronization, e.g.
recombination and fragmentation
mechanisms\cite{greco05,greco04,fries05}. On the other hand, if we
could get a definite $\Sigma^0/\Lambda$ yield ratio via some other
approaches, the information on the interactions between $\Lambda$'s
could then be obtained after the influence of $\Sigma^0/\Lambda$ was
removed.

\acknowledgements{We thank N. Xu and H. Huang for fruitful
discussions. We are grateful to J. Fu's kind helps. This work was
supported in part by the grants NSFC10447123 and 2004036221.}

\end{document}